\documentclass[aip,
 amsmath,amssymb,
 reprint,%
]{revtex4-1}

\usepackage{graphicx}
\usepackage{dcolumn}
\usepackage{bm}

\usepackage[utf8]{inputenc}
\usepackage[T1]{fontenc}
\usepackage{mathptmx}
\usepackage{etoolbox}

\usepackage[version=4]{mhchem}
\usepackage{chemfig}
\usepackage{chemmacros}
\usechemmodule{reactions}

\chemsetup[reactions]{
  before-tag = R ,
  tag-open = ( ,
  tag-close = )
}

\usepackage{floatrow}
\usepackage[caption=false,justification=justified, singlelinecheck=off]{subfig}
\newcommand{{\cbvn}}{C\textsubscript{B}V\textsubscript{N}}

\newcommand{\symmetry}[2]{$#1_{#2}$}
\definecolor{salmon}{RGB}{255,229,204}
\definecolor{lightblue}{RGB}{204,229,255}
\definecolor{lightgreen}{RGB}{229,255,204}

\usepackage{rotating}
\newlength\longest
\usepackage{nomencl}
\makenomenclature

\makeatletter
\def\@email#1#2{%
 \endgroup
 \patchcmd{\titleblock@produce}
  {\frontmatter@RRAPformat}
  {\frontmatter@RRAPformat{\produce@RRAP{*#1\href{mailto:#2}{#2}}}\frontmatter@RRAPformat}
  {}{}
}%
\makeatother

\begin{document}

\preprint{AIP/123-QED}

\title[Charging of quantum emitters in hexagonal boron nitride - graphene heterostructures due to electrostatic screening]{Charging of quantum emitters in hexagonal boron nitride - graphene heterostructures due to electrostatic screening}
\author{M. K. Prasad}
\affiliation{ 
School of Mathematics, Statistics and Physics, Newcastle University, Newcastle upon Tyne, NE1 7RU, United Kingdom
}
\affiliation{ 
Joint Quantum Centre (JQC) Durham-Newcastle, United Kingdom
}%

\author{J. P. Goss}%
 \email{jonathan.goss@ncl.ac.uk}
\affiliation{ 
School of Mathematics, Statistics and Physics, Newcastle University, Newcastle upon Tyne, NE1 7RU, United Kingdom
}

\author{J. D. Mar}%
 \email{jonathan.mar@ncl.ac.uk}
\affiliation{ 
School of Mathematics, Statistics and Physics, Newcastle University, Newcastle upon Tyne, NE1 7RU, United Kingdom
}%
\affiliation{ 
Joint Quantum Centre (JQC) Durham-Newcastle, United Kingdom
}%

\date{\today}

\begin{abstract}
Defect color centers in hexagonal boron nitride (hBN) have gained significant interest as single-photon emitters and spin qubits for applications in a wide range of quantum technologies. As the integration of these solid-state quantum emitters into electronic devices necessitates ctrical ntrol, it is essential to gain a deeper understanding of the mechanisms of charge control for these defect color centers in hBN/graphene heterostructures. In this Letter, we show that screening due to the encapsulation of hBN with graphene modifies the electrical levels of hBN, leading to charge transfer. Furthermore, we show that the charged defects have low-energy barriers for defect reorientation which can be overcome by moderate gate voltages. This study shows that accurate modeling of the charge state of the defect is necessary to be able to electrically control defects.
\end{abstract}

\maketitle



Defects in hBN have gained significant interest for their applications in quantum technologies, defect-based LEDs and spin-valve devices \cite{Tran2016,grzeszczyk2024electroluminescence,asshoff2018magnetoresistance,prasad2024hexagonal}. The architecture of such devices involves hBN being encapsulated in a heterostructure, such as by graphene \cite{Noh2018}. It was found that when graphene is interfaced with hBN, emitters were quenched due to charge and energy transfer \cite{Xu2020,Stewart2021}, hence it is clear that graphene has a potentially critical role in the control of defects in hBN via electrical fields or charge injection \cite{grzeszczyk2024electroluminescence,yu2022electrical,Xu2020,Stewart2021}. It is therefore necessary to study the impact of graphene layers on the electrical levels of defects in hBN, noting the modification of electrical levels also has an impact on the response of defects to external electric fields \cite{Noh2018,Scavuzzo2019}. For example, Stark effect experiments show that defects which possess an electric dipole perpendicular to the hBN plane show an inversion of dipole orientations or charging in the presence of strong electric fields parallel to the dipole orientation \cite{Noh2018}.

In our previous work it was shown that the location of the donor and acceptor levels of defects in hBN relative to the work function of graphene is a valuable predictor for charge transfer between hBN and graphene, and we demonstrated a method using the electronic density of states (DOS) to predict the degree of charge transfer \cite{prasad2023charge}. However, as the electrical levels of defects are sensitive to both strain and external environment, it is of interest to study the impact of graphene layers on the electrical levels of defects in hBN \cite{sensoy2017strain,aschauer2016interplay,deger2022lattice,Wang2019,Wang2020}. 

In this Letter, using the carbon--nitrogen-vacancy complex ({\cbvn}) as an illustrative example, we show that charge transfer occurs when hBN is encapsulated by graphene, although such a process is not considered to be energetically favorable based on calculations of the donor level in isolated monolayer hBN. We also show that due to strain there is an out-of-plane distortion of the defect and demonstrate that the barrier to the inversion of the direction of the dipole moment is significantly lowered due to charging. Our findings therefore contribute significantly to the understanding of electric field control via the Stark effect, electroluminescence and strain control of hBN quantum emitters.


Our DFT calculations were performed using the \textit{Ab Initio} Modelling PROgram\citep{jones1998identification} (AIMPRO) with periodic boundary conditions and the PBE-GGA exchange-correlation functional \citep{perdew1996generalised}. The atoms are modeled using norm-conserving separable pseudo potentials~\citep{hartwigsen1998relativistic}. Kohn-Sham eigenfunctions are represented with a basis of sets of independent $s$- and $p$-Gaussian orbitals ~\citep{goss-TAP-104-69}, with the addition of one (two) sets of $d$-Gaussian functions for C (B and N) atoms to account for polarization. The charge density is Fourier transformed using plane waves with an energy cutoff of 300\,Ha, leading to energies converged to better than 1{\,meV} with respect to this parameter. The Brillouin zone of the primitive structures were sampled using a $16\times16\times1$ Monkhorst-Pack scheme \citep{monkhorst1976special}; non-primitive cells employed sampling with the same or denser reciprocal space density. Structures were optimized by the conjugate-gradient method until the total energy changed by less than $10^{-5}$\,Ha, and forces are less than $10^{-4}$\,a.u. 

\begin{figure}[!ht]
        \centering
        \sidesubfloat[]{\label{fig: cbvn structure in hbn isolated}\includegraphics[width=0.35\textwidth]{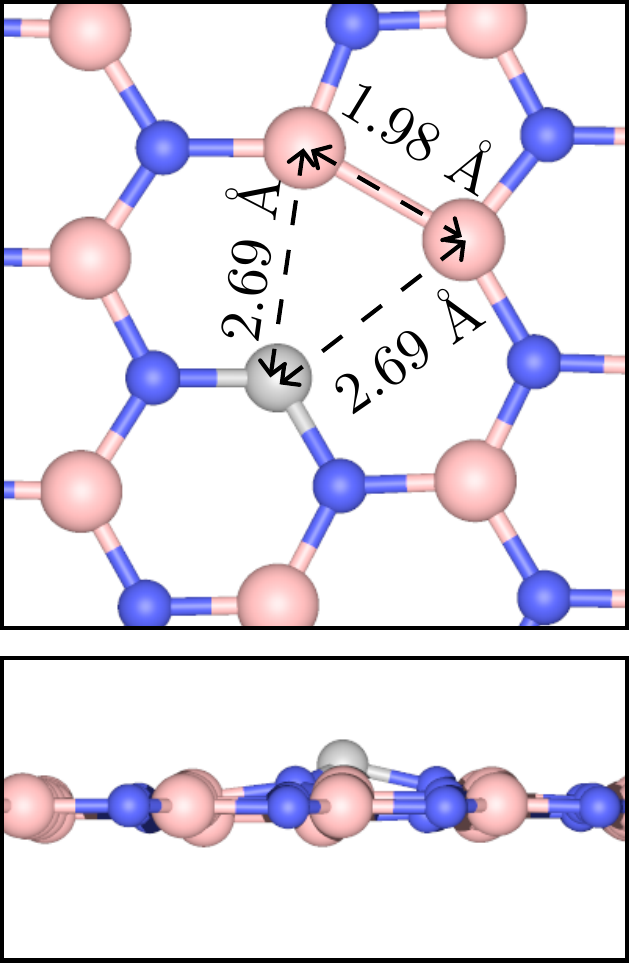}}\hspace{1em}
        \sidesubfloat[]{\label{fig: cbvn band structure in hbn isolated}\includegraphics[width=0.48\textwidth]{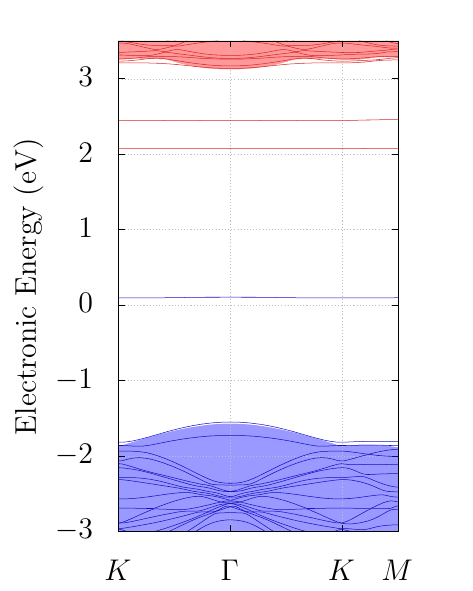}}
        \caption{\textbf{(a)} Schematic of {\cbvn} in its singlet ground-state, with pink, blue and gray spheres representing B, N and C, respectively. \textbf{(b)} Shows the corresponding band structure, where the shading indicates the filling of the corresponding defect-free hBN. Blue and red shading indicate occupied and empty bands, respectively. The zero in the energy scale is set to the Dirac point of the pristine graphene.}
        \label{fig: cbvn in hBN}
\end{figure}

The vacuum thickness was greater than four interlayer spacings of bulk hBN (Ref.\,\onlinecite{wang2017graphene}) and Van der Waals interactions were represented using the Grimme-D3 scheme~\citep{grimme2011density}. 
The formation energy of a defect, $X$, in charge state $q$ is
\begin{equation}\label{eqn: formation energy}
E_f(X,q) = E_\text{tot}(X,q) - E_\text{host} - \sum_i n_i \mu_i + q\left(\varepsilon_\text{VBM} + \varepsilon_F \right),
\end{equation}
where $E_\text{tot}(X,q)$ is the total energy of the defective system, $E_\text{host}$ is the total energy of pristine monolayer hBN of the same size, $n_i$ is the change in the number of atoms of species $i$ relative to pure hBN and $\mu_i$ is the chemical potential of the species $i$. The formation energies were calculated for N-rich conditions specified by $\mu_\text{N} +\mu_\text{B} = \mu_\text{hBN},$ where $\mu_\text{N}$ is half the total energy of an N$_2$ molecule and $\mu_\text{hBN}$ is the energy per formula unit of monolayer hBN. $\varepsilon_F$  is the electron chemical potential relative to the host valence band maximum (VBM), $\varepsilon_\text{VBM}$. To mitigate against the systematic errors due to periodic boundary conditions (PBC) in charged systems, we have adopted the uniform scaling of the cell sizes \cite{komsa2014charged,komsa2018erratum},
leading to an uncertainty of the order of $\pm0.1$\,eV in the formation energy \citep{komsa2018erratum}.

Quantification of charge transfer has been approached by integrating the DOS of the heterostructure from the Fermi level to the small band gap induced by the formation of the heterostructure \cite{prasad2023charge}. Calculations were performed with the net spin either constrained (fixed spin) or allowed to vary during the self-consistency process (free spin). For free spin calculations, the population of the spin channels was based upon Fermi-Dirac statistics with spin-up and spin-down channels having the same self-consistent electron chemical potential.

Nudged elastic band (NEB) calculations were performed using the climbing-image method \cite{henkelman2000climbing} for a $6\times6$ supercell, and eleven images between the initial and final states was found to sufficiently resolve the energy surface.

C\textsubscript{B}V\textsubscript{N} is formed when carbon substitutes for boron adjacent to a nitrogen vacancy. Interest in {\cbvn} has focused on its potential as a source of visible single-photon emission \cite{sajid2020vncb,mendelson2021identifying}, particularly as it is among a class of defects with an out-of-plane intrinsic electric dipole, rendering it accessible to Stark tuning via external electric fields \cite{Tawfik2017,Scavuzzo2019}. The ground state structure of the defect is a spin singlet, with \symmetry{C}{s} symmetry due to the out-of-plane movement of the carbon atom \cite{linderalv2021vibrational} (Fig.\,\ref{fig: cbvn structure in hbn isolated}) . The introduction of this defect leads to one (two) occupied (unoccupied) non-degenerate defect-levels in the band gap (Fig.\,\ref{fig: cbvn band structure in hbn isolated}).

\begin{figure}
    \centering
    \sidesubfloat[]{\label{fig: cbvn band structure in 1 hbn/gr ab}\includegraphics[width=0.42\textwidth]{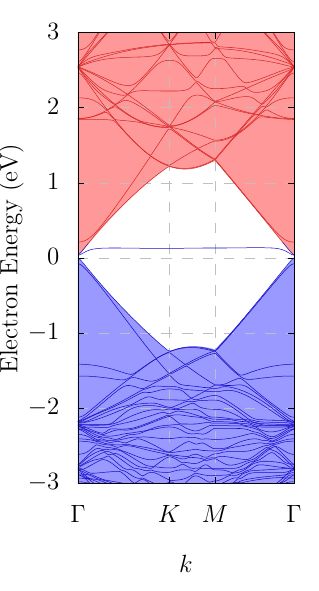}}\hspace{1em}
    \sidesubfloat[]{\label{fig: cbvn band structure in 2 hbn/gr ab}\includegraphics[width=0.42\textwidth]{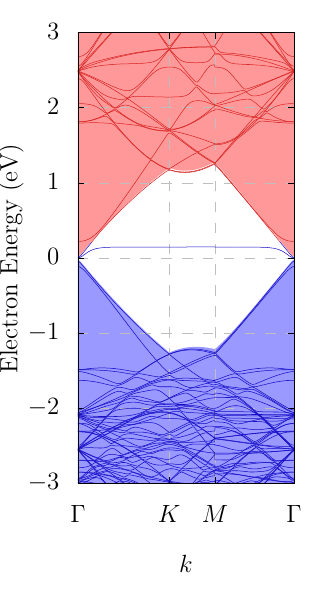}}\\
    \caption{Band structures of {\cbvn} in \textbf{(a)} hBN/Gr and \textbf{(b)} 2-hBN/Gr, with {\cbvn} in the hBN layer adjacent to Gr. The zero in the energy scale is set to Dirac point the corresponding the pristine heterostructure. Colors follow Fig.\,\ref{fig: cbvn in hBN}}
    \label{fig: band structure in hBN/Gr and 2-hBN/Gr}
\end{figure}

The defect when either positively and negatively charged forms a spin doublet, with $C_{s}$ symmetry geometries \cite{mendelson2021identifying}. Taking the average $z$ position of the boron and nitrogen atoms in hBN as the location of the plane, a small, 0.18\,\r{A}, out-of-plane distortion in the positively charged defect lowers the energy by $\sim2$\,meV. This has not been captured by existing studies in literature \,\cite{cheng2017paramagnetic}. Although it is a small difference, strain effects significantly enhance the stability of the distorted arrangement, as will be discussed later. The donor and acceptor levels of {\cbvn} were calculated at 5.1\,eV and 1.3\,eV below vacuum. Since the ionization energy exceeds the graphene work function (calculated at 4.3\,eV \cite{prasad2023charge,ziegler2011variations}), charge transfer is not expected. Free spin structural relaxation of {\cbvn} in hBN/Gr resulted in zero effective spin in the system, consistent with the defect remaining the in the neutral charge state as determined by explicit calculations of the degree of charge transfer.

A similar result was observed when an additional hBN layer was added to the system, resulting in bilayer hBN on graphene (2-hBN/Gr). No charge transfer was found and the spin-singlet state remained the ground state regardless of the layer location of the defect. This is reflected in the band structures of {\cbvn} in both hBN/Gr and 2-hBN/Gr, Fig.\,\ref{fig: band structure in hBN/Gr and 2-hBN/Gr}.

The impact of an additional layer of Gr on hBN/Gr to form Gr/hBN/Gr is now reviewed. Structural relaxation of the defect-free three-layer structure yields an in-plane lattice constant of 2.48\,\r{A}, representing a {+0.5\%} and {$-1.3$\%} strains in the graphene and hBN layers relative to the isolated monolayers.

\setlength{\fboxsep}{0pt}
\setlength{\fboxrule}{1pt}
\begin{figure}[!ht]
        \centering
        \sidesubfloat[]{\label{fig: cbvn singlet structure in gr/hbn/gr ab}\fbox{\includegraphics[width=0.4\textwidth]{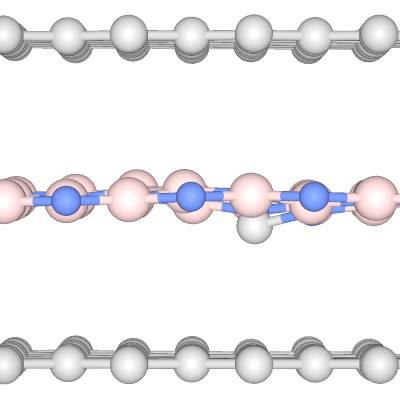}}}\hspace{1em}
        \sidesubfloat[]{\label{fig: cbvn doublet structure in gr/hbn/Gr ab}\fbox{\includegraphics[width=0.4\textwidth]{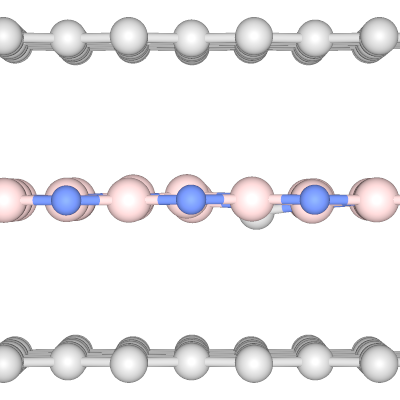}}}\\
        \sidesubfloat[]{\label{fig: cbvn singlet band structure plan in gr/hbn/gr}\includegraphics[width=0.31\textwidth]{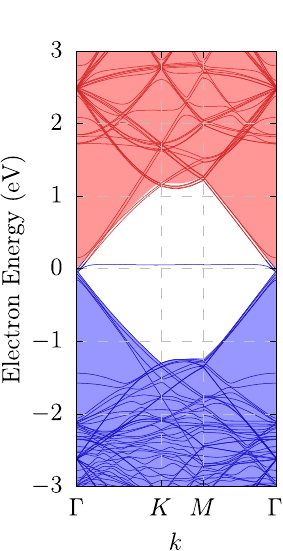}}
        \sidesubfloat[]{\label{fig: cbvn doublet band structure side up in gr/hbn/gr}\includegraphics[width=0.55\textwidth]{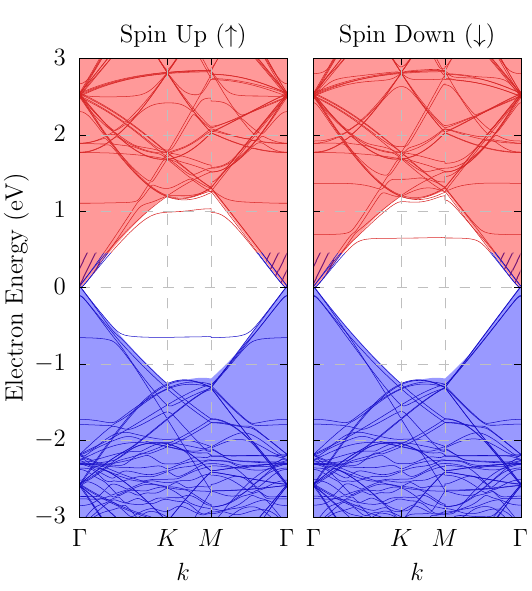}}
        \caption{The side view of {\cbvn} in Gr/hBN/Gr in showing the displacement of the carbon atom from the hBN plane in \textbf{(a)} the spin singlet state and \textbf{(b)} the spin doublet state (from free spin calculations). \textbf{(c)} and \textbf{(d)} show the corresponding band structures of the singlet and doublet configurations, respectively. The zero in the energy scale is set to the Dirac point of pristine Gr/hBN/Gr. The hatched shading reflects the occupation of the graphene states due to charge transfer from hBN. Colors follow Fig.\,\ref{fig: cbvn in hBN} and \ref{fig: band structure in hBN/Gr and 2-hBN/Gr}.}
        \label{fig: cbvn in Gr/hBN/Gr}
\end{figure}

Structural relaxation performed with free spin and fixed spin ($S=0$) calculations resulted in the carbon atom moving out-of-plane. The carbon atom moved 0.24\,nm during free spin relaxation from the hBN plane, significantly less than the 0.54\,nm displacement obtained for in the fixed spin-singlet relaxation (Fig.\,\ref{fig: cbvn singlet structure in gr/hbn/gr ab} and \ref{fig: cbvn doublet structure in gr/hbn/Gr ab}). This contrasts the observation in hBN/Gr and 2-hBN/Gr where the free spin and fixed spin calculations resulted in the same structure. The reason for the difference in the structure can be understood from the magnetic moment in the free-spin case, which is $\sim1$\,$\mu_B$, $\mu_B$ being the Bohr magneton. 
The calculation of charge transfer revealed that $\sim e$ transfers from {{\cbvn}} to graphene, resulting in a spin doublet. 
In our calculations, the self-consistently obtained spin doublet corresponding to the transfer of an electron to graphene, is only marginally stable, being just 20\,meV lower than the energy obtained for the fixed spin, representing an overall spin-singlet. 
The calculations were repeated in $8\times8$ and $10\times10$ supercells, and in both cases the spin doublet is lower in energy (by 99\,meV and 120\,meV, respectively).
We also established that the vacuum spacing does not impact the energy ordering of the $S=0$ and $S=\frac{1}{2}$ states, reflecting the absence of net polarization between the layers, as determined from the complete absence of any canceling electric field in the vacuum. 
As charge transfer occurred for Gr/hBN/Gr and not hBN/Gr, we suggest that the electrical levels of {\cbvn} are modified when encapsulated by graphene.

There are two significant features of interest in this Gr/hBN/Gr system. 
First, it was observed that charge transfer occurs only when hBN is encapsulated in graphene. 
Second, the geometry of {\cbvn} in the positive charged state in isolated hBN is nearly planar in geometry, a significant out-of-plane distortion occurs in the heterostructure.

\begin{figure}[!ht]
        \centering
        \includegraphics[width=0.88\textwidth]{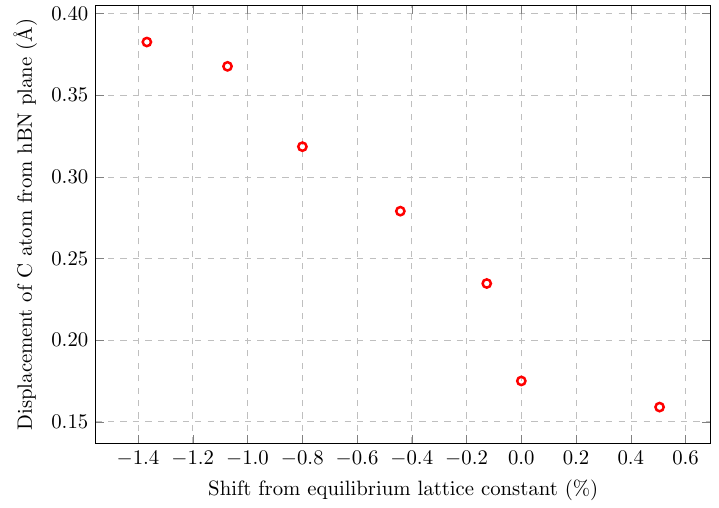}\\
        \caption{Displacement of C from the hBN layer in positively charged {\cbvn} in monolayer hBN  as a function of isotropic biaxial strain.}
        \label{fig: cbvn pos with strain}
\end{figure}

To explore the latter of these, the positively charged {\cbvn} in isolated monolayer hBN was optimised under different biaxial strains. 
Figure\,\ref{fig: cbvn pos with strain} shows that increasing compressive strain leads to an increase in the displacement of the carbon atom from the hBN plane. 
At approximately $-1.3\,\%$ strain, at which the lattice constant of hBN is equal to that of the heterostructure, the carbon atom is 0.38\,\r{A} from the hBN plane. At the same strain, the restriction of the carbon atom to the hBN plane results in a configuration 50\,meV lower in energy, highlighting that stabilisation due to distortion, although small in unstrained hBN, has physical significance.



\begin{figure}[!ht]
        \centering
        \includegraphics[width=\textwidth]{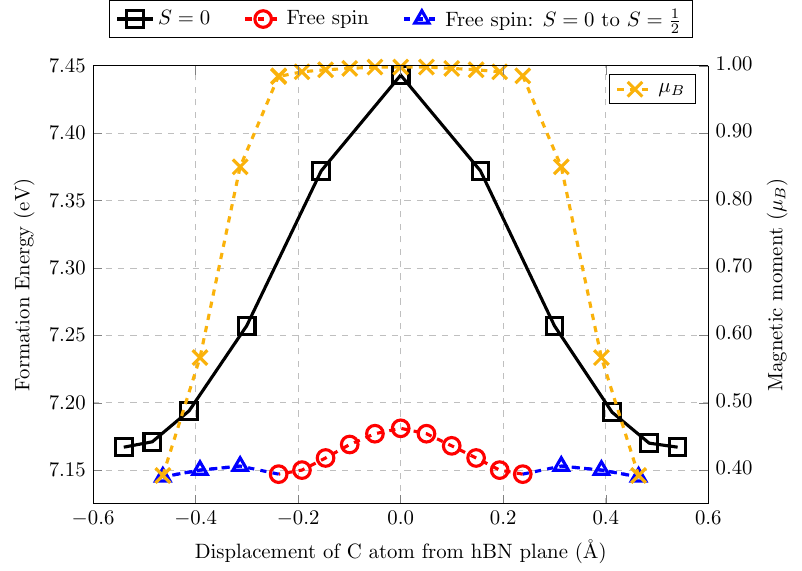}
        \caption{Energies for the NEB path between the equilibrium configurations of the spin singlet state (squares), doublet state (circles) and between the singlet and doublet states (triangles). Crosses show the variation in $\mu_B$ for the free spin case.}
        \label{fig: neb of gr/hbn/gr}
\end{figure}

As both the spin-singlet and doublet configurations were found in the Gr/hBN/Gr heterostructure, we sought to understand the barrier to move between the two states (noting that displacement of C above and below the hBN plane are equivalent by symmetry).
The barrier between the two directions of displacement was investigated for $S=0$ and $S=\frac{1}{2}$ states, the result of these NEB calculation being  presented in Fig.\,\ref{fig: neb of gr/hbn/gr}. 

The NEB data for the $6\times6$ supercell show the charge-transfer, non-zero effective spin case (circles and triangles in Fig.\,\ref{fig: neb of gr/hbn/gr}), is energetically favored throughout the inversion process. 
The energies for the largest displacement represented by the triangles are slightly lower than the corresponding limits for the squares, and we note that this energy difference converges to around 120{\,meV} for larger cells. 
The increase in the energy difference is attributed to the convergence of the effective spin from $0.98\,\mu_B$ to $1\,\mu_B$ with cell-size for the $8\times8$ and $10\times10$ cells, consistent with the convergence of the degree of charge transfer. 
For the data shown by circles the effective spin remains close to $\frac{1}{2}$, indicating that charge-transfer is constant through this range, and the out-of-plane displacement  stabilizes the defect. 
The barrier to move between the two orientations where charge-transfer occurs is 30\,meV, whereas for the fixed spin-singlet case the barrier is 280\,meV. 

\begin{figure}[!ht]
        \centering
        \includegraphics[width=0.88\textwidth]{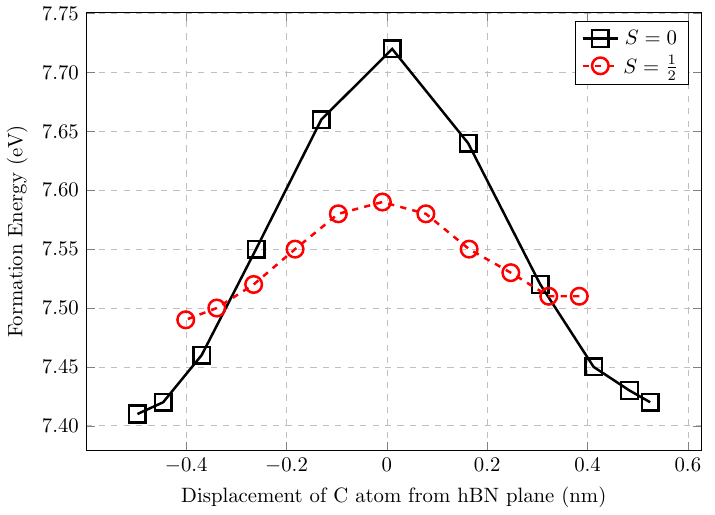}
        \caption{The variation of the formation energy of {\cbvn} along the energy path for $S=0$ (squares) and $S=\frac{1}{2}$ (circles) configurations in 2-hBN/Gr.}
        \label{fig: cbvn neb in 2 hbn/gr}
\end{figure}

For comparison with a system where no charge transfer is observed, the energy profile (Fig.~\ref{fig: cbvn neb in 2 hbn/gr}) for flipping the carbon orientations for the spin singlet and doublet states are shown for {\cbvn} in 2-hBN/Gr, where the calculations were performed with fixed spin. At the minima in the $S=\frac{1}{2}$ curve, the singlet configuration is lower in energy for the same displacement of the carbon atom. Hence, a repetition of the structural relaxation with free spin converges to the $S=0$ equilibrium configuration. Comparing 2-hBN/Gr with Gr/hBN/Gr (Fig.\,\ref{fig: cbvn neb in 2 hbn/gr} and \ref{fig: neb of gr/hbn/gr}), suggests that the donor level has been raised above the Dirac point due to the inclusion of the second graphene layer, leading to charge transfer.

We now comment on the implications for the observable optical properties based upon our findings. 
It is clear from Fig.\,\ref{fig: neb of gr/hbn/gr} and \ref{fig: cbvn neb in 2 hbn/gr} that the encapsulating environment has a significant effect on the {\cbvn} charge transition levels. 
Similar pheonomena were previously predicted in a study on the variation of the charge transition levels (CTLs) of hBN with additional hBN layers, \ce{SiO2} and diamond substrates \cite{Wang2019,Wang2020}. 
Here we find that additional Gr layers have a much greater effect on the CTLs than additional hBN layers. 
The change may be attributed to the difference in the electric permittivities of hBN and graphene, and hence the magnitude of the electrostatic screening of the defect. 
As such, unlike in 2-hBN/Gr and 1-hBN/Gr systems, the charge-transfer is predicted for Gr/hBN/Gr. 
This will be especially relevant in systems where multilayer hBN is sandwiched between graphene gates for the application of electric fields and charge injection \cite{Noh2018,Scavuzzo2019,yu2022electrical,white2022vdw}.

In monolayer hBN there is a small displacement carbon atom from the hBN plane for ({\cbvn})$^+$, and the effect is enhanced in the heterostructures. 
As illustrated by the data in Fig.\,\ref{fig: cbvn pos with strain}, displacement of carbon from the hBN plane is stabilized by increasing compressive biaxial strain. 
Therefore, under strain or via the influence of the substrate even the positively charged defect can be tuned by an electric field perpendicular to the hBN plane, as it possesses a component of the electric dipole moment parallel to the applied field. 
However, the low barrier between the two equilibria of ({\cbvn})$^+$ suggests that low gate voltages are sufficient for the atoms to overcome the barrier and flip to the site on the opposite side of the plane. 
This would invert the orientation of the dipole with respect to the field and subsequently the direction of the Stark shift. 
Studies on Stark tuning of hBN emitters have found `V' shaped variations of zero-phonon line with applied field, which supports the occurrence of such processes \cite{Noh2018}. 
It is also possible that some defects can possess charged configurations within the barrier of transition and act as metastable intermediate states for the transition between sites.

As the energy difference between the spin singlet and doublet configurations is $\sim120$\,meV, continued increase in the applied voltages would be likely to lead to a change in the charge state and turn the emitter dark. 
Even if the defect remains optically active after charge transfer, the change in the orientation of the electric dipole moment would manifest as an abrupt change in the polarization and degree of Stark shift. 
A low barrier to transition also increases the probability of tunneling of the carbon atom between the sites.



In this Letter we have shown that the encapsulation of hBN by graphene layers leads to dramatic modifications in the electrical levels. 
Using {\cbvn} as an illustrative example, we show that the donor level shifts toward the conduction band due to additional graphene layers, leading to charge transfer. NEB modeling also shows that the barrier to the flipping of the carbon atom across the hBN plane is relatively low and would likely to lead to a `V' shaped Stark shift, such as illustrated in Ref.\,\onlinecite{Noh2018}. 
It is therefore imperative that in determining viable color centers for single photon emitters that may be controlled by an external electric field, the correct charge state, strain and substrate are explicitly modeed. 
Also, the complex interplay between the dielectric environment and strain led to defect geometries that can differ significantly from the monolayer case and has implications on spectroscopic properties, such as the response to external electric fields and the degree of phonon coupling. 
The results of this work will therefore have a significant impact on the electric field, charge and strain control of quantum emitters in hBN for applications in quantum technologies.

\end{document}